\documentclass[prd,aps,showpacs,floatfix,twocolumn,superscriptaddress]{revtex4}

\def\rmd{{\rm d}}
\def\p{\partial}

\usepackage{amssymb,graphicx,color}

\begin{document}

\title{Finite differencing second order systems\\describing
  black hole spacetimes}

\author{Gioel Calabrese}

\affiliation{School of Mathematics, University of Southampton,
Southampton, SO17 1BJ, UK}

\date{\today}

\begin{abstract}
Keeping Einstein's equations in second order form can be appealing for
computational efficiency, because of the reduced number of variables
and constraints.  Stability issues emerge, however, which are not
present in first order formulations.
We show that a standard discretization of the second order ``shifted''
wave equation leads to an unstable semi-discrete scheme if the shift
parameter is too large.  This implies that discretizations obtained
using integrators such as Runge-Kutta, Crank-Nicholson, leap-frog are
unstable for any fixed value of the Courant factor.  We argue that
this situation arises in numerical relativity, particularly in
simulations of spacetimes containing black holes, and discuss several
ways of circumventing this problem.  We find that the first order
reduction in time based on ``ADM'' type variables is very effective.

\end{abstract}

\pacs{02.70.Bf, 04.25.Dm}

\maketitle


\section{Introduction}

In recent years there has been a growing interest in discretizing the
second order Einstein's equations, in the harmonic gauge or its
generalization, without reducing the system to first order form
\cite{Gar,SziWin,Pre}.  The reduction process requires the
introduction of auxiliary variables approximating first derivatives of
the fields and the introduction of additional constraints.  Whereas
there are clear advantages in keeping the system of equations in
second order form, including the fact that local well-posedness of the
continuum Cauchy problem has been shown \cite{Bru} and the expectation
that in general this would lead to smaller numerical errors
\cite{KPY}, we point out difficulties that can arise when a standard
discretization is used.

After analyzing a toy model problem that captures the essential
difficulty, and pointing out its relevance to numerical relativity, we
discuss different solutions to this problem.  The first order
reduction in time based on the introduction of ``ADM'' type variables
seems to be the most attractive of these solutions.

\section{The shifted wave equation}

We start with the wave equation in one spatial dimension,
$\phi_{\tilde t\tilde t} = \phi_{\tilde x\tilde x}$, and perform a
Galilean change of coordinates, $t = \tilde t$, $x = \tilde x-\beta
\tilde t$, where $\beta$ is a constant.  This leads to
\begin{equation}
\phi_{tt} = 2\beta \phi_{tx} + (1-\beta^2)\phi_{xx}\,, \label{Eq:shiftedwave}
\end{equation}
which we will refer to as the shifted wave equation.  By performing a
differential reduction to first order we see that the
characteristic variables and speeds are $\phi_t -\beta\phi_x \pm
\phi_x$, $\beta \pm 1$.  The variable $\phi$ is also a characteristic
variable, the speed of which is undetermined (it depends on the
details of the reduction and one can choose what one pleases).  The
initial value problem for this system is well-posed for any value of
$\beta$.  In fact, an energy estimate can be obtained by noting that
the quantity
\begin{equation}
\int \left((\phi_t-\beta\phi_x)^2 + \phi_x^2\right) \rmd x
\label{Eq:consenergy}
\end{equation}
is positive definite in $\phi_t$, $\phi_x$ and is conserved for any
$\beta$. 

We introduce the grid $x_j = jh$, where $h$ is the space step, and the
grid-function $\phi_j(t)$ approximating $\phi(t,x_j)$.  Leaving time
continuous, the standard second order accurate approximation of
Eq.~(\ref{Eq:shiftedwave}) is
\begin{equation}
\frac{d^2\phi_j}{dt^2} = 2\beta D_0\frac{d\phi_j}{dt} +
(1-\beta^2)D_+D_-\phi_j\,,\label{Eq:semiwave}
\end{equation}
where $hD_+u_j = u_{j+1}-u_j$, $hD_- u_j= u_j - u_{j-1}$ and $2D_0 =
D_++D_-$.  Consider the discrete quantity
\begin{equation}
(\phi_t,\phi_t)_h + 
(1-\beta^2)  (D_+\phi,D_+\phi)_h\,, \label{Eq:discenergy}
\end{equation}
where $(u,v)_h = \sum_j u_jv_jh$.  For $|\beta| < 1$ this expression
is positive definite in $\phi_t$, $D_+\phi$ and is
conserved\footnote{To show this one has to use the identities $D_+D_-
= D_-D_+$, $(u,D_0u)_h = 0$, $(u,D_\pm v)_h = - (D_{\mp} u, v)_h$, the
proofs of which can be found in \cite{GKO-Book}.  Note that the
continuum limit of expression (\ref{Eq:discenergy}) is not given by
expression (\ref{Eq:consenergy}).  Whereas the latter is equivalent to
$\int (\phi_t^2 + \phi_x^2)\rmd x$ for any value of $\beta$, the first
is only equivalent to it for $|\beta|<1$.}.  As in the continuum case,
the energy estimate follows.  The semi-discrete system is stable.

On the other hand, if $|\beta|>1$, there does not exist a positive
definite quantity from which one can derive a discrete energy
estimate.  A closer look at Eq.~(\ref{Eq:semiwave}) reveals that there
might be a problem with the highest frequency (and those nearby), due
to the fact that $D_0$ is unable to see it, $D_0 (-1)^j = 0$.
Consequently, at this frequency Eq.~(\ref{Eq:semiwave}) appears to be
elliptic.  It is not difficult to show that the semi-discrete problem
admits solutions that grow exponentially without bound in $h$.
Inserting $\phi_j(t) = e^{st}\bar\phi_j$ into Eq.~(\ref{Eq:semiwave})
we obtain
\[
\tilde s^2 \bar\phi_j = \tilde s \beta(\bar\phi_{j+1}-\bar\phi_{j-1}) +
(1-\beta^2) (\bar\phi_{j+1} - 2\bar\phi_j + \bar\phi_{j-1})\,,
\]
where $\tilde s = sh$.  For $\bar\phi_j = (-1)^j$, we get
$\tilde s^2 = 4(\beta^2-1)$.  Hence, the grid-function 
\begin{equation}
\phi_j(t) =
e^{2\sqrt{\beta^2-1}t/h}(-1)^j\label{Eq:growingmode}
\end{equation}
is a solution of Eq.~(\ref{Eq:semiwave}), the growth of which cannot
be bounded independently of $h$.  Notice that this analysis also
applies to the first order in time, second order in space system
\begin{eqnarray}
\frac{d\phi_j}{dt} &=& T\,,\label{Eq:semiwave2}\\
\frac{dT_j}{dt} &=& 2\beta D_0 T_j + 
(1-\beta^2)D_+D_-\phi_j\,.\nonumber
\end{eqnarray}
In particular, this shows that schemes such as the forward Euler,
backward Euler, Runge-Kutta, Crank-Nicholson and leap-frog methods
applied to either Eq.~(\ref{Eq:semiwave}) or system
(\ref{Eq:semiwave2}) are unstable if $|\beta|>1$.  The scheme is also
unstable for $|\beta| = 1$.  However, in this case the instability is
less severe (the system admits linearly growing frequency dependent
solutions).

A toy model problem similar to the shifted wave equation was
considered by Alcubierre and Schutz \cite{AlcSch}, who proved
instability for an implicit scheme and proposed using causal
differencing \cite{SeiSue,Schetal,GunWal,LehHuqGar} to eliminate the
instability.  Our semi-discrete analysis leads to a more general
result, namely that the instability is due to the spatial
discretization and does not depend on the time integration.
Furthermore, it is important to realize that this type of instability
does not appear in fully first order systems.  In these cases one can
handle high values of the characteristic speeds by choosing a
sufficiently small Courant factor (and possibly adding artificial
dissipation in the variable coefficient case).  Whenever causal
differencing has been applied to first order systems, it has not
brought substantial improvements \cite{Schetal,GunWal}.

Before discussing how we propose to fix this problem, we show how it
can arise in discretizations of second order systems of Einstein's
equations.  For concreteness, we consider formulations having
principal part determined by a wave operator of the form
$g^{\mu\nu}\p_{\mu}\p_{\nu}$, where $g^{\mu\nu}$ is the inverse
4-metric of spacetime.  Precisely this operator appears in the
(generalized) harmonic gauge \cite{Wal}.  We keep the system in second
order form and use the standard spatial discretization.  Assuming that
the coordinates are chosen such that the $t={\rm const}.$ slices are
space-like, i.e., $g^{tt}<0$, one can expect the instability to arise
whenever the spatial coordinates are such that an $x^i={\rm const.}$
hyper-surface is also space-like, i.e., $g^{ii}<0$ (no sum).  Again,
to the highest grid frequency this problem appears to be elliptic.
Interestingly, the last condition, $g^{ii}<0$, is a requirement for
excision, a technique often used in numerical relativity to eliminate
the black hole singularity from the computational domain.  This shows
that when discretizing second order systems describing spacetimes
containing black holes, one has to ponder over the discretization.

Another instance in which this type of instability can arise is when
rigidly co-rotating coordinates are used.  These coordinates are
introduced to attempt to keep a binary black hole system at a fixed
coordinate location \cite{Bruetal,Dueetal}.  At large distances the
semi-discrete wave operator effectively becomes elliptic for the
highest frequencies.

We now go back to the model equation (\ref{Eq:semiwave}) and discuss
several methods to overcome the instability that occurs for
$|\beta|>1$, without reducing the spatial derivative.

{\em Method 1:} We know that the addition of artificial dissipation
can sometimes stabilize otherwise unstable schemes.  If we modify
system (\ref{Eq:semiwave2}) as follows
\begin{eqnarray}
\frac{d\phi_j}{dt} &=& T_j - \sigma h^3 (D_+D_-)^2 \phi_j\,,\label{Eq:meth1}\\
\frac{dT_j}{dt} &=& 2\beta D_0T_j + (1-\beta^2)D_+D_-\phi_j - \sigma h^3
(D_+D_-)^2T_j\,, \nonumber
\end{eqnarray}
we see that the von Neumann condition, which is only a necessary
condition for stability, is satisfied for sufficiently large values of
the dissipation parameter (for example $\sigma \gtrsim 0.385$ for
$|\beta| = 2$).  However, such a great amount of dissipation demands
high resolution to prevent excessive damping and requires a rather
small Courant factor.  For fourth order Runge-Kutta (4RK) in the
$|\beta|=2$ case we need $k/h \lesssim 0.289$, where $k$ is the time
step.

{\em Method 2:} Perhaps the simplest solution is to replace the one
sided operators $D_{\pm}$ in Eq.~(\ref{Eq:semiwave}) with the centered
one, $D_0$.  This amounts to discretizing the second spatial
derivatives with the $D_0^2$ operator instead of $D_+D_-$, as
suggested in \cite{Bonetal,BabSziWin}, leading to a scheme with a five
point stencil instead of three.  With such discretization the discrete
version of (\ref{Eq:consenergy}), with the replacement $\p_x \to D_0$,
is conserved and a von Neumann stability analysis gives a Courant limit of
$\sqrt{8}/(1+|\beta|)$ for 4RK.

At first glance this method appears to be very effective.  It
suppresses the exponentially growing mode (\ref{Eq:growingmode}) and
it allows for a rather large time step.  However, the fact that $D_0$
is blind to the highest frequency means that the discrete conserved
quantity is unable to capture it and, as discussed in greater detail
in \cite{CHHR}, the method is not robust in the sense that a
perturbation of the equation by lower order terms can trigger
(exponentially growing) numerical instabilities.  Although it is
possible that artificial dissipation may cure this problem, this needs
to be explored.  Whatever the case may be, a five point stencil is
likely to unduly complicate the treatment of boundaries.


{\em Method 3:} Another alternative is to rewrite
Eq.~(\ref{Eq:shiftedwave}) as
\begin{eqnarray}
\p_t \phi &=& \beta \p_x \phi + \Pi\,, \label{Eq:adm1cont}\\
\p_t \Pi &=& \beta \p_x \Pi +\p^2_x \phi\,, \nonumber
\end{eqnarray}
where we have introduced the variable $\Pi = \p_t \phi - \beta\p_x
\phi$.  The standard second order accurate discretization now gives
\begin{eqnarray}
\left(\frac{d}{dt} - \beta D_0\right) \phi_j &=& \Pi_j\,,
\label{Eq:adm1}\\ \left(\frac{d}{dt} - \beta D_0\right) \Pi_j &=&
D_+D_-\phi_j\,. \nonumber
\end{eqnarray}
Note that in terms of the original second order system this spatial
discretization corresponds to
\begin{equation}
\frac{d^2\phi_j}{dt^2} - 2\beta D_0 \frac{d\phi_j}{dt} +
\beta^2D_0^2\phi_j = D_+D_-\phi_j\,, \label{Eq:Wmixed}
\end{equation}
which has a five point stencil.  Incidentally, for large $\beta$ it is
not possible to construct a centered, second order accurate, three
point stencil, stable approximation of the second order equation
(\ref{Eq:shiftedwave}), without performing a first order reduction in
time.  System (\ref{Eq:adm1}) is stable for any value of $\beta$, as
it conserves the discrete quantity
\begin{equation}
(\Pi,\Pi)_h + (D_+\phi,D_+\phi)_h\,.
\end{equation}
With 4RK and for large $\beta$ it has a Courant limit comparable to
that of method 2.  In particular, for $|\beta|=2$ we get the condition
$k/h \lesssim 0.803$.  Furthermore, the numerical speeds of
propagation associated with system (\ref{Eq:adm1}) are closer to the
exact ones than those of method 2.

Both the continuum system and approximation (\ref{Eq:adm1}) are
non-dissipative.  They admit a conserved energy.  If the advective
terms (the terms multiplied by $\beta$) in the semi-discrete system
(\ref{Eq:adm1}) are approximated with second order accurate one-sided
operators $D_+(1-\frac{h}{2}D_+)$ rather than $D_0$, assuming
$\beta>1$, we are trading a conservative scheme with one which is
dissipative.  Although in the variable coefficient case this may be
effective in obtaining stability, in this particular case, with 4RK
and $\beta =2$, the scheme requires $k/h \lesssim 0.332$, which is
less than half what is needed by the centered approximation.

Finally, we point out that in the fourth order accurate case, $\p_x
\to D_0(1-\frac{1}{6}h^2D_+D_-)$, $\p_x^2 \to
D_+D_-(1-\frac{1}{12}h^2D_+D_-)$, the results of this paper remain
qualitatively unchanged.  For $|\beta| > 1$ the standard
discretization of (\ref{Eq:shiftedwave}) is unstable, unless copious
amount of artificial dissipation is added to the scheme\footnote{In
this case with 4RK and sixth order dissipation one needs $\sigma
\gtrsim 0.081$ and $k/h \lesssim 0.303$ for $|\beta| = 2$.
Surprisingly, when going from second to fourth order accuracy, this
system allows for a larger Courant factor.}, whereas the
discretization of (\ref{Eq:adm1cont}) is stable for any value of the
shift parameter.

\section{Conclusion}

Our analysis demonstrates that when using formulations which are
second order in space one has to exercise caution, even in the absence
of boundaries.  We find that the instability discussed in
\cite{AlcSch}, which motivated the introduction of causal
differencing, is due to the mixing of the $D_0$ and $D_+D_-$ operators
in the spatial discretization and therefore only appears with second
order in space systems.  The fact that it arises already at the
semi-discrete level, as in Eq.~(\ref{Eq:semiwave}), shows that no time
integrator can fix it, not even implicit ones.  One can expect such
instability to arise in numerical relativity simulations based on
standard spatial discretization of fully second order systems near
black holes and at large distances from the center of a rigidly
rotating coordinate system.

We believe that a simple and effective method of eliminating the
instability consists in rewriting the system in ``ADM'' form before
discretizing it, as in (\ref{Eq:adm1}).  When this is done, the
resulting approximation, which is still centered, is stable for any
value of the parameter $\beta$.  It is straightforward to prove
stability using the discrete energy method and higher order accurate
schemes can be easily constructed.  Interestingly, the structure of
system (\ref{Eq:adm1cont}) is similar to commonly used first order in
time, second order in space formulations of Einstein's equations, such
as the Baumgarte-Shapiro-Shibata-Nakamura system \cite{SN,BS}.


\begin{acknowledgments}
We wish to thank Carsten Gundlach, Ian Hawke, Ian Hinder, Luis Lehner,
David Neilsen, Oscar Reula, and Olivier Sarbach, for helpful
discussions and suggestions.  This research was supported by a Marie
Curie Intra-European Fellowship within the 6th European Community
Framework Program.
\end{acknowledgments}



\begin{thebibliography}{99}

\bibitem{Gar} D.~Garfinkle,
Phys.\ Rev.\ D {\bf 65}, 044029 (2002).

\bibitem{SziWin} B.~Szilagyi and J.~Winicour,
Phys.\ Rev.\ D {\bf 68}, 041501 (2003).

\bibitem{Pre} F.~Pretorius, gr-qc/0407110.

\bibitem{Bru} Y.~Bruhat, ``The Cauchy Problem'', in {\it Gravitation:
an introduction to current research}, edited by L.~Witten, John Wiley
and Sons (1962).

\bibitem{KPY} H.~Kreiss, N.~Petersson, and J.~Ystr\"om, SIAM
J.~Numer.~Anal.~{\bf 40}, 1940-1967 (2002).

\bibitem{AlcSch} M.~Alcubierre and B.~Schutz, J.~Comput.~Phys.~{\bf
112}, 44 (1994).


\bibitem{SeiSue} E.~Seidel and W.M.~Suen, Phys.~Rev.~Lett.~{\bf 69},
1845 (1992).

\bibitem{Schetal} M.~Scheel, T.~Baumgarte, G.~Cook, S.~Shapiro, and
S.~Teukolsky, Phys.~Rev.~D {\bf 56}, 6320 (1997).


\bibitem{GunWal} C.~Gundlach and P.~Walker, Class.~Quantum Grav.~{\bf
16}, 991 (1999).

\bibitem{LehHuqGar} L.~Lehner, M.~Huq, and D.~Garrison, Phys.~Rev.~D
{\bf 62}, 084016 (2000).

\bibitem{Wal} R.~Wald, {\it General Relativity} (University of Chicago
Press, Chicago, 1984).

\bibitem{Bruetal}
B.~Br\"ugmann, W.~Tichy, and N.~Jansen, Phys.~Rev.~Lett.~{\bf 92},
211101 (2004)

\bibitem{Dueetal}
M.D.~Duez, P.~Marronetti, S.L.~Shapiro, and T.W.~Baumgarte,
Phys.~Rev.~D {\bf 67}, 024004 (2003).

\bibitem{Bonetal} C.~Bona, T.~Ledvinka, C.~Palenzuela, and M.~\v Z\'
a\v cek, Phys.~Rev.~D {\bf 69}, 064036 (2004).

\bibitem{BabSziWin} M.~Babiuc, B.~Szilagyi, and J.~Winicour,
gr-qc/0404092.

\bibitem{CHHR} G.~Calabrese, I.~Hinder, S.~Husa, K.~Roszkowski, in
preparation.

\bibitem{SN} M.~Shibata and T.~Nakamura, Phys.~Rev.~D {\bf 52}, 5428
(1995).

\bibitem{BS} T.~Baumgarte and S.~Shapiro, Phys.~Rev.~D {\bf 59},
024007 (1999).

\bibitem{GKO-Book} B.~Gustafsson, H.~Kreiss, and J.~Oliger, {\em Time
dependent problems and difference methods} (John Wiley \& Sons, New
York, 1995).

\end{thebibliography}
\end{document}